\documentclass[a4paper,11pt]{article}
\usepackage{pos}

\usepackage{acronym}
\usepackage[capitalize]{cleveref}
\usepackage{tabularx,booktabs}

\newcommand{\citere}[1]{Ref.~\cite{#1}}
\newcommand{\citeres}[1]{Refs.~\cite{#1}}

\acrodef{ibp}[IBP]{integration-by-parts}

\title{Towards the next Kira release}

\author*[a,b]{Fabian Lange}
\author[c]{Johann Usovitsch}
\author[d]{Zihao Wu}

\affiliation[a]{Physik-Institut, Universit\"at Z\"urich,\\
  Winterthurerstrasse 190, 8057 Z\"urich, Switzerland}
\affiliation[b]{Paul Scherrer Institut,\\
  5232 Villigen PSI, Switzerland}
\affiliation[c]{Theoretical Physics Department, CERN,\\
  1211 Geneva, Switzerland}
\affiliation[d]{School of Fundamental Physics and Mathematical Science, Hangzhou Institute of Advanced Study, UCAS, \\
Hangzhou 310024, China}

\emailAdd{fabian.lange@physik.uzh.ch}
\emailAdd{johann.usovitsch@cern.ch}
\emailAdd{wuzihao@mail.ustc.edu.cn}

\abstract{
The reduction of Feynman integrals to a basis of master integrals plays a crucial role for many high-precision calculations and \texttt{Kira} is one of the leading tools for this task.
In these proceedings we discuss some of the new features and improvements currently being developed for the next release.
}

\FullConference{Loops and Legs in Quantum Field Theory (LL2024)\\
 14-19, April, 2024\\
Wittenberg, Germany\\}

\begin{document}
\maketitle

\section{Introduction}

Feynman integrals play a central role in precision calculations, both in the Standard Model and beyond.
A direct integration of all integrals of an amplitude, nowadays often many tens or hundreds of thousands, is usually unfeasible.
The default strategy in modern precision calculations is thus to first reduce the integrals to a much smaller set of master integrals employing \ac{ibp} identities~\cite{Tkachov:1981wb,Chetyrkin:1981qh} and the Laporta algorithm~\cite{Laporta:2000dsw}.
This strategy is implemented in the public programs \texttt{AIR}~\cite{Anastasiou:2004vj}, \texttt{FIRE}~\cite{Smirnov:2008iw,Smirnov:2013dia,Smirnov:2014hma,Smirnov:2019qkx,Smirnov:2023yhb}, \texttt{Reduze}~\cite{Studerus:2009ye,vonManteuffel:2012np}, \texttt{Kira}~\cite{Maierhofer:2017gsa,Klappert:2020nbg}, \textsc{FiniteFlow}~\cite{Peraro:2019svx} in combination with \texttt{LiteRed}~\cite{Lee:2012cn,Lee:2013mka}, and most recently \texttt{Blade}~\cite{Guan:2024byi}.

After an overview of \ac{ibp} reductions and \texttt{Kira} in \cref{sec:overview}, we discuss some of the features and improvements that we are developing for the next release in \cref{sec:features}.
We highlight the performance improvements with some benchmarks in \cref{sec:benchmarks} and discuss our findings and future directions in \cref{sec:discussion}.

\section{An overview over integration-by-parts reductions and Kira}
\label{sec:overview}

In general, a scalar $L$-loop Feynman integral can be written as
\begin{equation}
  T_j \equiv T(a_1,\dots,a_N) = \int \left(\prod\limits_{i=1}^L \mathrm{d}^d\ell_i\right) \frac{1}{P_1^{a_1} P_2^{a_2} \cdots P_N^{a_N}} ,
  \label{eq:Feynman-integral}
\end{equation}
where
\begin{equation}
  P_j=q_j^2-m_j^2, \quad j=1,\dots,N
  \label{eq:propagators}
\end{equation}
are inverse Feynman propagators.
The momenta $q_j$ are linear combinations of the loop momenta $\ell_i$, $i=1,\dots,L$, and external momenta $p_k$, $k=1,\dots,E$ for $E+1$ external legs (or $E=0$ for vacuum integrals), and $m_j$ are the propagator masses.
The $a_j$ are the (integer) propagator powers.
The set of inverse propagators must be complete and independent in the sense that every scalar product of momenta can be uniquely expressed as a linear combination of the $P_j$, squared masses $m_j^2$, and external kinematical invariants.
The number of propagators is thus $N=\frac{L}{2}(L+2 E+1)$ including auxiliary propagators that only appear with $a_j\le 0$.

The Feynman integrals of \cref{eq:Feynman-integral} are in general not independent.
In \citeres{Tkachov:1981wb,Chetyrkin:1981qh} it was found that they are related by so-called \acf{ibp} identities
\begin{equation}
  \int \left(\prod\limits_{i=1}^L \mathrm{d}^d\ell_i\right) \frac{\partial}{\partial \ell^\mu_i} \frac{q^\mu_j}{P_1^{a_1} P_2^{a_2} \cdots P_N^{a_N}} = 0 ,
\end{equation}
where $q^\mu_j$ is either a loop momentum, an external momentum, or a linear combination thereof.
In addition to these \ac{ibp} identities, there also exist so-called Lorentz-invariance identities~\cite{Gehrmann:1999as}
\begin{equation}
  \sum_{i=1}^E \left( p_i^\nu \frac{\partial}{\partial p_{i\mu}} - p_i^\mu \frac{\partial}{\partial p_{i\nu}} \right) T(a_1,\dots,a_N) = 0 .
\end{equation}
They do not provide additional information, but it was found that they often facilitate the reduction process.
Finally, there are symmetry relations between Feynman integrals which are most often searched for by comparing graph polynomials with Pak's algorithm~\cite{Pak:2011xt}.
All these relations provide linear equations of the form
\begin{equation}
  0 = \sum_i c_i(\{s_j, m_k\}) T(a_{1,i},\dots,a_{N,i})
  \label{eq:linear-relation}
\end{equation}
between different Feynman integrals, where the coefficients $c_i(\{s_j, m_k\})$ are rational functions of the kinematic invariants $s_j$ and the propagator masses $m_k$.
The relations can be solved to express the Feynman integrals in terms of a basis of master integrals.
It was even shown that the number of master integrals for the standard Feynman integrals of \cref{eq:Feynman-integral} is finite~\cite{Smirnov:2010hn}.

Nowadays, the most prominent solution strategy is the Laporta algorithm~\cite{Laporta:2000dsw}:
The linear relations of \cref{eq:linear-relation} are generated for different values of the propagator powers $a_j$ resulting in a system of equations between specific Feynman integrals.
Each choice of $\{a_j\}$ is called a \emph{seed} and this generation process is also known as \emph{seeding}.
The system can then be solved with Gauss-type elimination algorithms.

This requires an ordering of the integrals.
In \texttt{Kira} we first assign the integrals to so-called topologies based on their respective sets of propagators, i.e.\ their momenta and masses.
We assign a unique integer number to each topology.
Secondly, we assign each integral to a sector
\begin{equation}
  S = \sum\limits_{j=1}^N 2^{j-1}\,\theta(a_j - \tfrac{1}{2}),
  \label{eq:sector}
\end{equation}
where $\theta(x)$ is the Heaviside step function.
A sector $S$ is called subsector of another sector $S'$ (with propagators powers $a'_j$) if $S<S'$ and $a_j\le a'_j$ for all $j=1,\dots,N$.
We denote as top-level sectors those sectors which are not subsectors of other sectors that contain Feynman integrals occurring in the reduction problem at hand.
For the discussions in the following sections it is beneficial to use the big-endian binary notation instead of the sector number defined by \cref{eq:sector}, i.e.\ the sector $S=5$ is represented by b$1010$ where each number represents one bit of the number $5$.
In this notation it is immediately visible that b$1010$ is a subsector of b$1110$.

Furthermore, it is useful to define the number of propagators with positive powers,
\begin{equation}
  t = \sum_{j=1}^N \theta(a_j - \tfrac{1}{2}),
  \label{eq:def-t}
\end{equation}
and
\begin{equation}
  r = \sum_{j=1}^N a_j \theta(a_j - \tfrac{1}{2}),\qquad
  s = -\sum_{j=1}^N a_j\theta(\tfrac{1}{2} - a_j),\qquad
  d = \sum_{j=1}^N (a_j - 1) \theta(a_j - \tfrac{1}{2})
\end{equation}
denoting the sum of all positive powers, the negative sum of all negative powers, and the sum of positive powers larger than $1$, respectively.
These concepts allow us to order the integrals in \texttt{Kira}.
Secondly, they are used to limit the seeds $\{a_j\}$ for which equations are generated.
We generate the equations only for those sets $\{a_j\}$ for which $r\le r_{\max}$, $s\le s_{\max}$, and $d\le d_{\max}$, where $r_{\max}$, $s_{\max}$, and $d_{\max}$ are input provided by the user.

The workflow in \texttt{Kira} is split into two components: first generating the system of equations and then solving it.
We offer two different strategies for the latter.
First, the system can be solved analytically using the computer algebra system \texttt{Fermat}~\cite{Fermat} for the rational function arithmetic.
Secondly, the system can be solved using modular arithmetic over finite fields~\cite{Kauers:2008zz,Kant:2013vta,vonManteuffel:2014ixa,Peraro:2016wsq}.
The variables, i.e.\ kinematic invariants and masses, are replaced by random integer numbers and all arithmetic operations are performed over prime fields.
Choosing the largest $63$-bit primes allows us to perform all operations on native data types and to avoid large intermediate expressions.
The final result can then be interpolated and reconstructed by \emph{probing} the system sufficiently many times for different random choices of the variables.
We employ the library \texttt{FireFly} for this task~\cite{Klappert:2019emp,Klappert:2020aqs}.

The main improvements in the upcoming version of \texttt{Kira} discussed in \cref{sec:features} concern the generation of the system of equations.
Let us thus outline the current algorithm in more detail:
\begin{enumerate}
  \item Identify trivial sectors and discard all integrals in these sectors.
  \item Identify symmetries between sectors including those of other integral families.
  Drop sectors which can identically be mapped away.
  The remaining symmetries give rise to additional equations to be generated.
  \item For every remaining sector starting from the simplest one, generate all equations within the specifications provided by the user:
  \begin{enumerate}
    \item Generate one equation.
    \item Check if it is linearly independent by inserting all previously generated equations and solving them numerically using modular arithmetic~\cite{Kant:2013vta}.
    \item Keep it if it is linearly independent, otherwise drop it.
  \end{enumerate}
  \item After all equations were generated, solve the system for the target integrals specified by the user with modular arithmetic and drop all equations which do not contribute to their solutions.
\end{enumerate}
At the end of this procedure \texttt{Kira} has generated a system of linearly independent equations specifically trimmed to suffice to reduce the target integrals to master integrals.

\section{Improvements and new features under development}
\label{sec:features}

In this section we discuss some of the improvements and new features which are currently under development.

\subsection{Internal reordering of propagators}

\begin{figure}[ht]
  \centering
  \includegraphics[trim = 36 538 36 36, scale = 0.4]{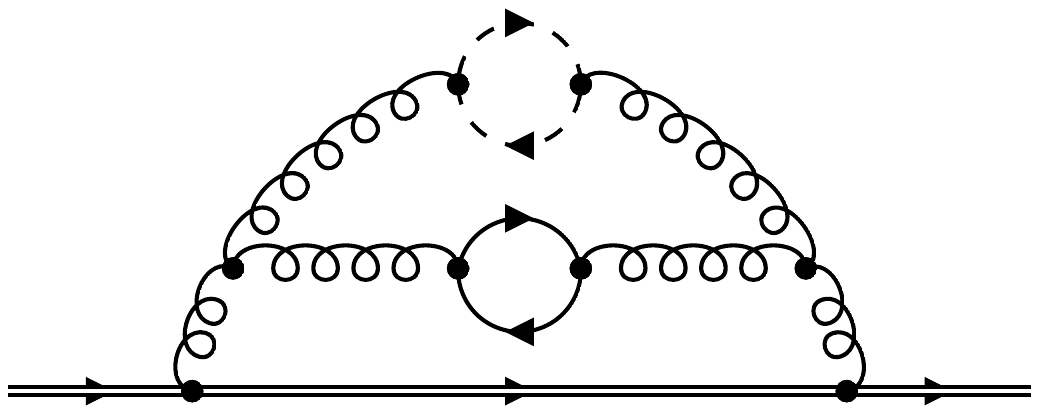}
  \caption{Four-loop propagator topology with two masses.}
  \label{fig:four-loop-propagator}
\end{figure}

It is well known that the definition of the propagators in \cref{eq:propagators} influences the performance of the reduction.
Not only different choices for the momenta have an impact, simply reordering them can make a difference, especially when selecting equations before the actual reduction as in \texttt{Kira}.
This can be demonstrated with the four-loop propagator type topology depicted in \cref{fig:four-loop-propagator}.
We reduce all integrals appearing in the quark self energy amplitude for the relation between the quark mass renormalization constants in the $\overline{\text{MS}}$ and the on-shell schemes~\cite{Fael:2021kyg}.
We compare propagator order $1$,
\begin{equation}
  \begin{split}
    &p_1^2 - m_1^2,\ (p_1 - q)^2,\ \underline{(p_2 + p_3)^2},\ (p_1 - p_3 + p_4 - q)^2,\ (p_3 - p_4)^2,\ \underline{(p_3 + q)^2},\ \underline{(p_2 + p_4)^2}, \\
    &p_2^2 - m_2^2,\ (-p_1 + p_2 + p_3 - p_4 + q)^2 - m_2^2,\ p_3^2, p_4^2,\ \underline{(p_1 + p_3)^2},\ \underline{(p_1 + p_4)^2},\ \underline{(p_1 + p_2)^2} ,
  \end{split}
  \label{eq:propagator-order-1}
\end{equation}
which was automatically chosen when generating the amplitude in \citere{Fael:2021kyg}, with propagator order $2$,
\begin{equation}
  \begin{split}
    &p_3^2,\ p_4^2,\ p_1^2 - m_1^2,\ p_2^2 - m_2^2,\ (p_1 - q)^2,\ (p_3 - p_4)^2,\ (p_1 - p_3 + p_4 - q)^2,\\
    &(-p_1 + p_2 + p_3 - p_4 + q)^2 - m_2^2,\ \underline{(p_2 + p_3)^2},\ \underline{(p_3 + q)^2},\ \underline{(p_2 + p_4)^2},\ \underline{(p_1 + p_3)^2},\\
    &\underline{(p_1 + p_4)^2},\ \underline{(p_1 + p_2)^2} ,
  \end{split}
  \label{eq:propagator-order-2}
\end{equation}
which follows the guidelines outlined in \citere{Maierhofer:2018gpa}:
order by the number of momenta and put massless propagators first.
The underlined entries denote auxiliary propagators which only appear in the numerator and should be placed at the end.
When comparing the systems of equations generated with \texttt{Kira\ 2.3} for the two orders in \cref{tab:propagator-order}, we see that the system generated with order 2 contains $10$\,\% less equations, $14$\,\% less terms, and can be solved $28$\,\% faster than the system generated with order 1.
\begin{table}[ht]
  \begin{center}
    \caption{Comparison of the generated and selected systems of equations for the two propagator orders defined in \cref{eq:propagator-order-1,eq:propagator-order-2} of the four-loop propagator topology depicted in \cref{fig:four-loop-propagator}.}
    \label{tab:propagator-order}
    \begin{tabular}{c|c|c}
      \toprule
      & order 1 & order 2 \\
      \midrule
      \# of equations & $3\,737\,521$ & $3\,362\,446$ \\
      \# of terms & $47\,122\,072$ & $40\,326\,035$ \\
      Probe time & $57$\,s & $41$\,s \\
      \bottomrule
    \end{tabular}
  \end{center}
\end{table}

Since version \texttt{2.3} \texttt{Kira} offers the possibility to reorder the propagators internally independent of the order specified in the integral family definition, i.e.\ the propagator order in the input and output files differs from the order \texttt{Kira} uses internally to perform the reduction.
The ordering can either be chosen manually or be adjusted automatically according to four different ordering schemes.
The one outlined above usually provides the best results.
With the next release of \texttt{Kira} we will automatically turn on this ordering scheme by default.

\subsection{Improved seeding}
\label{ssec:seeding}

The choice of which equations should be generated for the reduction has severe impact on the performance of the reduction.
\texttt{Kira} seeds conservatively and generates all equations within the bounds specified by the user to reduce to the minimal basis of master integrals.
As outlined in \cref{sec:overview}, irrelevant equations are filtered out through a selection procedure.
However, for integrals with many loops or high values of $r$, $s$, and $d$ the combinatorics of the seeds easily gets out of hand and already makes the generation of the system of equations unfeasible, running both into runtime and memory limits.
We thus revised the seeding process in \texttt{Kira} and identified three areas for improvement: seeding of sectors related by symmetries, seeding of sectors containing preferred master integrals, and seeding of subsectors.
We address the three areas in the following.

In the \texttt{Kira 2.3}, sectors which are related by symmetries to other sectors are seeded with the same values for $r_{\max}$, $s_{\max}$, and $d_{\max}$.
Similarly, sectors which contain one of the preferred master integrals are seeded with the maximum values of $r_{\max}$, $s_{\max}$, and $d_{\max}$ from all sectors.
Both choices were made to ensure to reduce to the minimal basis, but turn out to be too conservative and negatively impacting performance.
In the upcoming release those sectors will be seeded with the default values for subsectors.

Currently the values for $r_{\max}$, $s_{\max}$, and $d_{\max}$ set for a sector are propagated down to all subsectors.
For example, if we are interested in the integral $T(1,1,1,1,1,1,1,0,-4)$, the typical choice is to seed the sector b$111111100$ with $r_{\max} = 7$ and $s_{\max} = 4$.
This then propagates to the lower sectors and sector b$111111000$ is automatically seeded with $r_{\max} = 7$, $s_{\max} = 4$, and $d_{\max} = 1$ and sector b$111000000$ with $r_{\max} = 7$, $s_{\max} = 4$, and $d_{\max} = 4$, respectively.
The increase in $d_{\max}$ can already be restricted by manually setting $d_{\max}$, but we previously argued against this to prevent finding to many master integrals.
With the new release we revise this recommendation and encourage setting $d_{\max}$.
However, keeping $s_{\max}$ constant when descending to the subsectors is even more problematic because the combinatorics of distributing irreducible scalar products in lower sectors grows quickly.
As it turns out, most of these seeds are actually not required to reduce integrals with positive $s$ in higher sectors and one can decrease $s_{\max}$ in lower sectors.
With the upcoming release we provide an option to restrict $s$ in lower sectors by $s \leq (t - l + 1)$, where $t$ was defined in \cref{eq:def-t} and $l$ can be chosen by the user.
This was first described in \citere{Driesse:2024xad}.
The same observation was also made by the developers of \texttt{Blade}~\cite{Guan:2024byi} and \texttt{FIRE}~\cite{Bern:2024adl} and is already available in the former code.

\subsection{Improved integral selection}

\begin{figure}
  \centering
  \includegraphics[scale = 0.7]{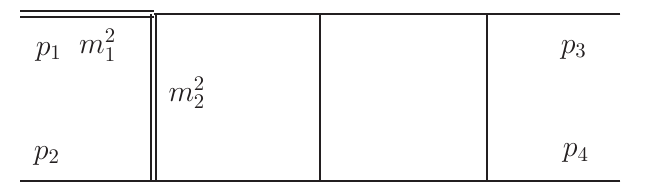}
  \caption{The planar double box \texttt{topo7} contributing, e.g., to single-top production.}
  \label{fig:topo7}
\end{figure}

The algorithm to select relevant equations in \texttt{Kira 2.3} is not optimal.
One weakness can easily be illustrated when reducing all integrals in the top-level sector b$111111100$ of the topology \texttt{topo7} depicted in \cref{fig:topo7} with exactly $s=5$.
First, we seed with $s_{\max}=5$ and, secondly, with $s_{\max}=6$, but in both cases select the equations for the same set of integrals.
As can be seen in \cref{tab:selection}, the number of selected equations and with it the number of terms as well as the time to solve the system increases.
\begin{table}
  \begin{center}
    \caption{Comparison of the selected systems for \texttt{topo7} when seeded with different values of $s_{\max}$.
    The improved seeding discussed in \cref{ssec:seeding} has been disabled as much as possible without major code modifications.}
    \label{tab:selection}
    \begin{tabular}{c|c|c|c|c}
      \toprule
      & \texttt{2.3}, $s_{\max}=5$ & \texttt{2.3}, $s_{\max}=6$ & \texttt{dev}, $s_{\max}=5$ & \texttt{dev}, $s_{\max}=6$ \\
      \midrule
      \# of selected equations & $62\,514$ & $85\,119$ & $34\,908$ & $34\,908$ \\
      \# of terms & $461\,336$ & $659\,803$ & $245\,975$ & $245\,975$ \\
      Probe time & $0.16$\,s & $0.24$\,s & $0.084$\,s & $0.084$\,s \\
      \bottomrule
    \end{tabular}
  \end{center}
\end{table}
This can be explained by the fact that \texttt{Kira} generates and selects sector by sector.
Thus, it prefers high seeds of low sectors over low seeds of higher sectors.
However, higher seeds most often result in more complicated equations.

Secondly, we identified another problem: \texttt{Kira} selects equations which do not contribute to the final result because they cancel in intermediate steps.
This can be illustrated with the toy system
\begin{equation}
  \begin{aligned}
    (\mathrm{i}) \quad 0=&\,x[6]+b_1\,x[2] , \\
    (\mathrm{ii}) \quad 0=&\,x[6]+c_1\,x[1] , \\
    (\mathrm{iii}) \quad 0=&\,x[5]+a_1\,x[2]+a_2\,x[4]+a_3 x[3] , \\
    (\mathrm{iv}) \quad 0=&\,x[5]+a_1\,x[2] ,
  \end{aligned}
  \label{eq:select-toy-initial}
\end{equation}
where $x[i]$ represents a Feynman integral and the Roman number in front denotes the equation number.
We now solve this system using Gaussian elimination.
After the forward elimination the system reads
\begin{equation}
  \begin{aligned}
    (\mathrm{i}) \quad 0=&\,b_1\,x[2]-c_1\,x[1] \quad &\{\mathrm{ii}\}& , \\
    (\mathrm{ii}) \quad 0=&\,x[6]+c_1\,x[1] \quad &\{\}& , \\
    (\mathrm{iii}) \quad 0=&\,a_2\,x[4]+a_3\,x[3] \quad &\{\mathrm{iv}\}& , \\
    (\mathrm{iv}) \quad 0=&\,x[5]+a_1\,x[2] \quad &\{\}& ,
  \end{aligned}
  \label{eq:select-toy-forward}
\end{equation}
where the numbers in the curly brackets denote which equations have been inserted.
After the back substitution we obtain
\begin{equation}
  \begin{aligned}
    (\mathrm{i}) \quad 0=&\,b_1\,x[2]-c_1\,x[1] \quad &\{\mathrm{ii}\}& , \\
    (\mathrm{ii}) \quad 0=&\,x[6]+c_1\,x[1]\, \quad &\{\}& , \\
    (\mathrm{iii}) \quad 0=&\,a_2\,x[4]+a_3\,x[3] \quad &\{\mathrm{iv}\}& , \\
    (\mathrm{iv}) \quad 0=&\,x[5]+a_1\,c_1/b_1\,x[1] \quad &\{\mathrm{i}, \mathrm{ii}\}& .
  \end{aligned}
  \label{eq:select-toy-final}
\end{equation}
\texttt{Kira 2.3} selects equations after the back substitution using the information in the curly brackets of \cref{eq:select-toy-final}.
If we are interested in the integral $x[4]$, it thus selects equation $(\mathrm{iii})$ because it provides the solution for $x[4]$, then equation $(\mathrm{iv})$ which was inserted into equation $(\mathrm{iii})$, and finally equations $(\mathrm{i})$ and $(\mathrm{ii})$ which were inserted into equation $(\mathrm{iv})$.
However, the latter two do not provide any information: it is easy to spot in \cref{eq:select-toy-initial} that equation $(\mathrm{iii})$ contains equation $(\mathrm{iv})$ as a subequation.
This subequation thus vanishes after inserting equation $(\mathrm{iv})$ only and the information of equations $(\mathrm{i})$ and $(\mathrm{ii})$ is not needed.
We call this phenomenon \emph{hidden zero} which in general can take more complicated forms.

So far we were not able to develop a systematic algorithm to detect all of the hidden zeros.
For the upcoming release we decided to instead rely on heuristics:
We select the equations for the target integrals after the forward elimination only and check if they contain all information to reduce to the basis of master integrals.
All non-reduced integrals are then added to the target integrals and we repeat the process until all target integrals are reduced to master integrals.

Returning to \cref{tab:selection}, we see that this improved selection algorithm addresses both problems described in this section: The development version of \texttt{Kira} selects the same equations even if seeding with higher values of $s_{\max}$ and less equations overall, indicating that we got rid of some hidden zeros.
Note that the improved seeding discussed in \cref{ssec:seeding} has been disabled as much as possible without major code modifications, but a thorough study of the improved selection alone is left for the future.

\subsection{Symbolic propagator powers}

\begin{figure}
  \centering
  \includegraphics[scale = 0.7]{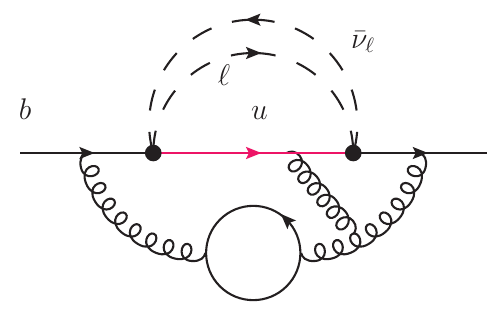}
  \caption{Diagram contributing to the $\alpha_{\mathrm{s}}^3$ corrections of the semileptonic $b \to u$ decay.}
  \label{fig:b2u-decay}
\end{figure}

With the upcoming version of \texttt{Kira} it will be possible to perform reductions with symbolic propagator powers by marking the associated propagators.
One application are the $\alpha_{\mathrm{s}}^3$ corrections to the semileptonic $b \to u$ decay considered in \citere{Fael:2023tcv}.
In the sample diagram shown in \cref{fig:b2u-decay} the lepton-neutrino loop does not receive QCD corrections and can simply be integrated out using
\begin{equation}
  \int \mathrm{d}^d p \frac{p^{\mu_1} \dots p^{\mu_N}}{(-p^2) [-(p-q)^2]} =
  \frac{\mathrm{i} \pi^{2-\epsilon}}{(-q^2)^{\epsilon}}
  \times \sum_{i=0}^{[N/2]}
  f(\epsilon,i,N) \left(\frac{q^2}{2} \right)^i \{[g]^i[q]^{N-2i}\}^{\mu_1 \dots \mu_N} ,
\end{equation}
where the propagator $-q^2$ is now raised to the power $\epsilon$.
Therefore, this trick reduces the five-loop reduction to a four-loop one at the price of one symbolic propagator power which can be handled by the upcoming version of \texttt{Kira}.

\section{Benchmarks}
\label{sec:benchmarks}

In this section we show two benchmarks to compare the current development version of \texttt{Kira} to the most recent release version \texttt{2.3}.

\subsection{Can one tune Kira 2.3 already?}

\begin{figure}
  \centering
  \includegraphics[width=0.3\textwidth, angle=-90]{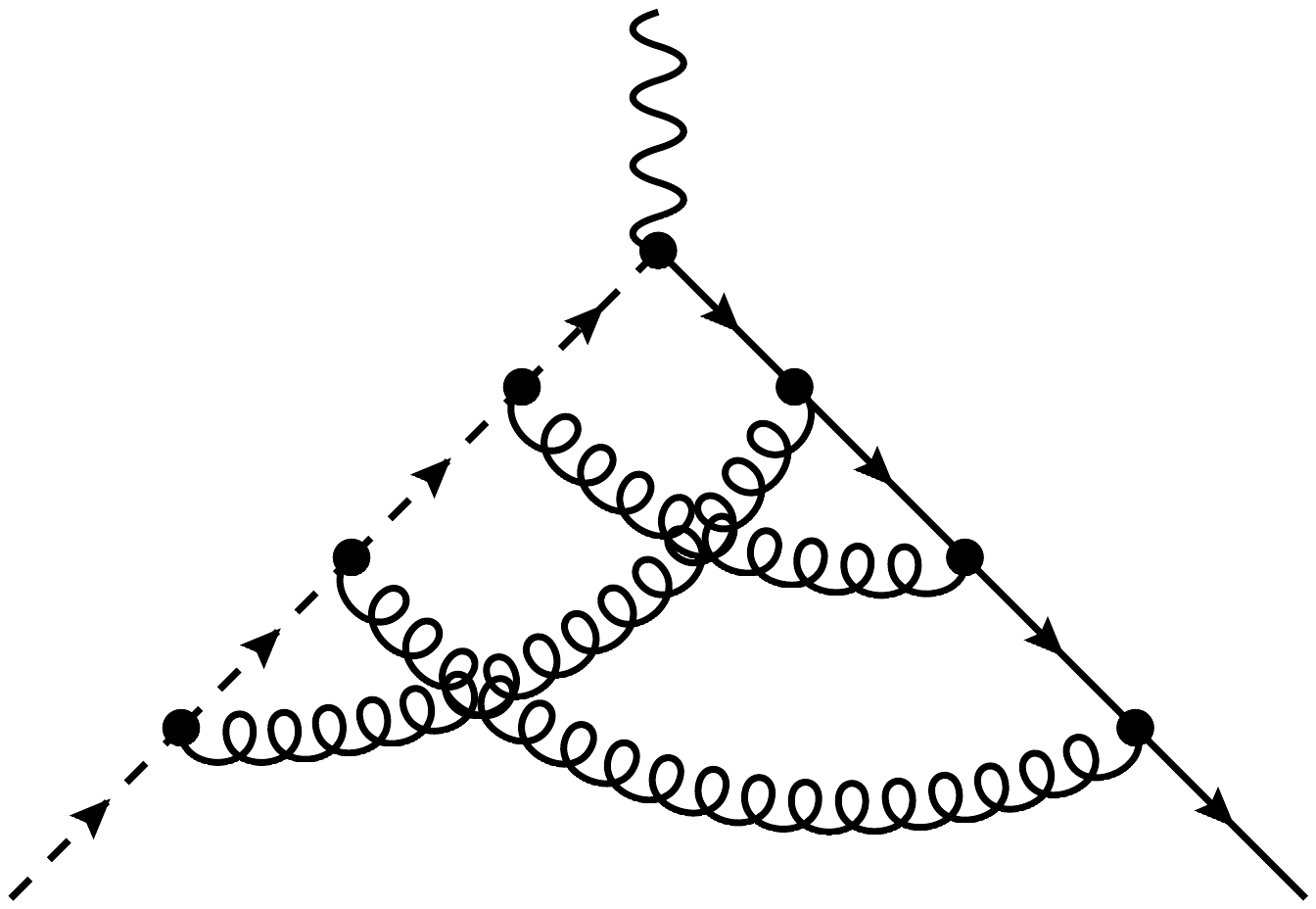}
  \caption{Topology contributing to the heavy-to-light quark form factors at the three-loop level.}
  \label{fig:heavy-to-light}
\end{figure}

Our first example is the vertex topology shown in \cref{fig:heavy-to-light} appearing in the heavy-to-light quark form factors at the three-loop level~\cite{Fael:2024vko}.
We want to reduce all integrals of the amplitude in $R_\xi$ gauge.
The curious reader might ask if it is already possible to achieve some of the improvements discussed in \cref{sec:features} by carefully tuning the input parameters in \texttt{Kira 2.3}, because they are just updated recommendations, e.g.\ setting $d_{\max}$.
We tried this to the best of our abilities, and failed as evident in \cref{tab:heavy-to-light}.
\begin{table}
  \begin{center}
    \caption{Comparison of the systems of equations for the topology shown in \cref{fig:heavy-to-light} with the recommended settings of \texttt{Kira 2.3}, an attempt to tune \texttt{Kira 2.3}, and the development version.}
    \label{tab:heavy-to-light}
    \begin{tabular}{c|c|c|c}
      \toprule
      & \texttt{Kira 2.3} & \texttt{Kira 2.3}, tuned & \texttt{Kira dev} \\
      \midrule
      \# of generated equations & $304\,821\,183$ & $198\,405\,903$& $3\,779\,262$ \\
      \# of selected equations & $21\,666\,745$ & $24\,708\,949$ & $828\,010$ \\
      \# of terms & $828\,060\,148$ & $583\,976\,288$ & $12\,516\,161$ \\
      Time to generate and select & $96\,359$\,s & $59\,579$\,s & $1\,328$\,s \\
      Memory to generate and select & $194$\,GiB & $129$\,GiB & $3.2$\,GiB \\
      Probe time & $2\,255$\,s & $3\,726$\,s & $19$\,s \\
      \bottomrule
    \end{tabular}
  \end{center}
\end{table}
While it is already possible to generate less equations compared to the previously recommended strategy, unfortunately this increases the number of selected equations and the time to solve the system significantly.
On the other hand, with the current development version we achieve a success on all fronts:
Compared to \texttt{Kira 2.3} with recommended settings, we generate $81$ times less equations and select $26$ times less equations which are overall simpler and contain $66$ times less terms.
The generation and selection speed up by a factor $73$ and require $61$ times less memory.
Finally, the time to solve the system numerically over a finite field reduces by a factor of $119$.

\subsection{Double pentagon}
\label{ssec:double-pentagon}

\begin{figure}
  \centering
  \includegraphics[scale = 0.7]{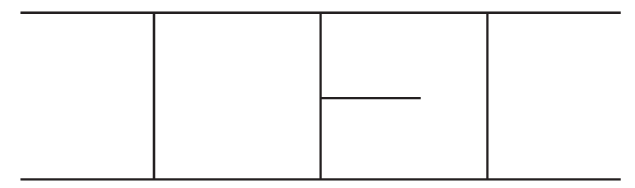}
  \caption{Massless double pentagon topology.}
  \label{fig:double-pentagon}
\end{figure}

As the second benchmark we choose the massless double pentagon topology shown in \cref{fig:double-pentagon}.
We reduce all integrals in the top-level sector b$11111111000$ with $s=5$.
Again we observe impressive improvements shown in \cref{tab:double-pentagon}:
\begin{table}
  \begin{center}
    \caption{Comparison of the systems of equations for the double pentagon topology shown in \cref{fig:double-pentagon} with the recommended settings of \texttt{Kira 2.3} and the development version. In the last column we reduce the latter system with \textsc{Ratracer}.}
    \label{tab:double-pentagon}
    \begin{tabular}{c|c|c|c}
      \toprule
      & \texttt{Kira 2.3} & \texttt{Kira dev} & \texttt{Kira dev} $+$ \textsc{Ratracer} \\
      \midrule
      \# of generated equations & $16\,872\,564$ & $76\,045$ & - \\
      \# of selected equations & $1\,157\,381$ & $41\,998$ & - \\
      \# of terms & $23\,053\,485$ & $734\,833$ & - \\
      Time to generate and select & $1\,259$\,s & $14$\,s & - \\
      Memory to generate and select & $30$\,GiB & $1.9$\,GiB & - \\
      Probe time & $38$\,s & $1.2$\,s & $0.12$\,s \\
      \bottomrule
    \end{tabular}
  \end{center}
\end{table}
We generate $222$ times less equations, select $28$ times less equations, and the system contains $31$ times less terms.
This speeds up the generation and selection by a factor $90$ and requires $16$ times less memory.
The probe time decreases by a factor of $32$.

Additionally, we use the program \textsc{Ratracer}~\cite{Magerya:2022hvj} to reduce the system generated by the development version.
It records the trace of operations performed on the input values and, thus, gets rid of the overhead associated to solving the system.
This way the probe time is sped up by another factor of $10$ which puts it well below $1$\,s, fast enough to sample the phase space as for example done in \citere{Agarwal:2024jyq}.

\section{Discussion}
\label{sec:discussion}

In these proceedings we presented some of the improvements and new features that we are developing for the next release of the Feynman integral reduction program \texttt{Kira}.
Besides the support for symbolic propagator powers, we focused on improvements to the algorithms to seed and select equations.
Our benchmarks show a significant speed-up and relaxation of the memory requirements by one or two orders of magnitude for the selected examples.
We already successfully applied the new version of \texttt{Kira} to three vastly different projects~\cite{Fael:2023tcv,Driesse:2024xad,Fael:2024vko}.

However, further investigation is in order:
So far, we only quantified the numbers for the generation of the system and the reduction using modular arithmetics over finite fields, neglecting the CPU time for the interpolation and reconstruction algorithms.
While this is negligible for small rational functions with only a few variables, it might dominate the runtime for complicated rational functions.
In this case there will be lesser improvements to the performance.
Thus, both technical and algorithmic improvements in \texttt{FireFly}~\cite{Klappert:2019emp,Klappert:2020aqs} might be necessary, e.g.\ following the idea of \citere{Chawdhry:2023yyx}.
On the other hand it might be worthwhile to sample the phase space for $2 \to 3$ processes instead of interpolating the reduction tables, as for example pursued in \citere{Agarwal:2024jyq}.
As shown \cref{ssec:double-pentagon}, the performance of the new version of \texttt{Kira} looks promising for this strategy, especially in combination with \textsc{Ratracer}~\cite{Magerya:2022hvj} or potentially the strategy presented in \citere{Liu:2023cgs}.

Furthermore, we did not investigate yet how much the improvements enhance the algebraic reduction with \texttt{Fermat}~\cite{Fermat}.
In this context it will also be interesting to study how other computer algebra systems might improve the performance~\cite{Mokrov:2023vva,Smirnov:2023yhb}.

Finally, we are looking into extending the symmetry finder algorithms following the ideas presented in \citere{Wu:2024paw} which will allow us to map more sectors away and reduce the number of master integrals.

\acknowledgments

The work of F.L.~was supported by the Swiss National Science Foundation (SNSF) under contract \href{https://data.snf.ch/grants/grant/211209}{TMSGI2\_211209}.
Some of the Feynman diagrams were drawn with the help of Axodraw~\cite{Vermaseren:1994je} and JaxoDraw~\cite{Binosi:2003yf} and some with the help of \texttt{FeynGame}~\cite{Harlander:2020cyh}.

\bibliographystyle{JHEP}
\bibliography{bib}

\providecommand{\href}[2]{#2}\begingroup\raggedright\begin{thebibliography}{10}

\bibitem{Tkachov:1981wb}
F.V.~Tkachov, \emph{{A theorem on analytical calculability of 4-loop
  renormalization group functions}},
  \href{https://doi.org/10.1016/0370-2693(81)90288-4}{\emph{Phys. Lett. B}
  {\bfseries 100} (1981) 65}.

\bibitem{Chetyrkin:1981qh}
K.G.~Chetyrkin and F.V.~Tkachov, \emph{{Integration by parts: The algorithm to
  calculate $\beta$-functions in 4 loops}},
  \href{https://doi.org/10.1016/0550-3213(81)90199-1}{\emph{Nucl. Phys. B}
  {\bfseries 192} (1981) 159}.

\bibitem{Laporta:2000dsw}
S.~Laporta, \emph{{High-precision calculation of multiloop Feynman integrals by
  difference equations}},
  \href{https://doi.org/10.1142/S0217751X00002159}{\emph{Int. J. Mod. Phys. A}
  {\bfseries 15} (2000) 5087}
  [\href{https://arxiv.org/abs/hep-ph/0102033}{{\ttfamily hep-ph/0102033}}].

\bibitem{Anastasiou:2004vj}
C.~Anastasiou and A.~Lazopoulos, \emph{{Automatic integral reduction for higher
  order perturbative calculations}},
  \href{https://doi.org/10.1088/1126-6708/2004/07/046}{\emph{JHEP} {\bfseries
  07} (2004) 046} [\href{https://arxiv.org/abs/hep-ph/0404258}{{\ttfamily
  hep-ph/0404258}}].

\bibitem{Smirnov:2008iw}
A.V.~Smirnov, \emph{{Algorithm FIRE -- Feynman Integral REduction}},
  \href{https://doi.org/10.1088/1126-6708/2008/10/107}{\emph{JHEP} {\bfseries
  10} (2008) 107} [\href{https://arxiv.org/abs/0807.3243}{{\ttfamily
  0807.3243}}].

\bibitem{Smirnov:2013dia}
A.V.~Smirnov and V.A.~Smirnov, \emph{{FIRE4, LiteRed and accompanying tools to
  solve integration by parts relations}},
  \href{https://doi.org/10.1016/j.cpc.2013.06.016}{\emph{Comput. Phys. Commun.}
  {\bfseries 184} (2013) 2820}
  [\href{https://arxiv.org/abs/1302.5885}{{\ttfamily 1302.5885}}].

\bibitem{Smirnov:2014hma}
A.V.~Smirnov, \emph{{FIRE5: A C++ implementation of Feynman Integral
  REduction}}, \href{https://doi.org/10.1016/j.cpc.2014.11.024}{\emph{Comput.
  Phys. Commun.} {\bfseries 189} (2015) 182}
  [\href{https://arxiv.org/abs/1408.2372}{{\ttfamily 1408.2372}}].

\bibitem{Smirnov:2019qkx}
A.V.~Smirnov and F.S.~Chukharev, \emph{{FIRE6: Feynman Integral REduction with
  modular arithmetic}},
  \href{https://doi.org/10.1016/j.cpc.2019.106877}{\emph{Comput. Phys. Commun.}
  {\bfseries 247} (2020) 106877}
  [\href{https://arxiv.org/abs/1901.07808}{{\ttfamily 1901.07808}}].

\bibitem{Smirnov:2023yhb}
A.V.~Smirnov and M.~Zeng, \emph{{FIRE 6.5: Feynman integral reduction with new
  simplification library}},
  \href{https://doi.org/10.1016/j.cpc.2024.109261}{\emph{Comput. Phys. Commun.}
  {\bfseries 302} (2024) 109261}
  [\href{https://arxiv.org/abs/2311.02370}{{\ttfamily 2311.02370}}].

\bibitem{Studerus:2009ye}
C.~Studerus, \emph{{Reduze -- Feynman integral reduction in C++}},
  \href{https://doi.org/10.1016/j.cpc.2010.03.012}{\emph{Comput. Phys. Commun.}
  {\bfseries 181} (2010) 1293}
  [\href{https://arxiv.org/abs/0912.2546}{{\ttfamily 0912.2546}}].

\bibitem{vonManteuffel:2012np}
A.~von Manteuffel and C.~Studerus, \emph{{Reduze 2 - Distributed Feynman
  Integral Reduction}},  \href{https://arxiv.org/abs/1201.4330}{{\ttfamily
  1201.4330}}.

\bibitem{Maierhofer:2017gsa}
P.~Maierh\"ofer, J.~Usovitsch and P.~Uwer, \emph{{Kira\textemdash{}A Feynman
  integral reduction program}},
  \href{https://doi.org/10.1016/j.cpc.2018.04.012}{\emph{Comput. Phys. Commun.}
  {\bfseries 230} (2018) 99}
  [\href{https://arxiv.org/abs/1705.05610}{{\ttfamily 1705.05610}}].

\bibitem{Klappert:2020nbg}
J.~Klappert, F.~Lange, P.~Maierh\"ofer and J.~Usovitsch, \emph{{Integral
  reduction with Kira 2.0 and finite field methods}},
  \href{https://doi.org/10.1016/j.cpc.2021.108024}{\emph{Comput. Phys. Commun.}
  {\bfseries 266} (2021) 108024}
  [\href{https://arxiv.org/abs/2008.06494}{{\ttfamily 2008.06494}}].

\bibitem{Peraro:2019svx}
T.~Peraro, \emph{{\textsc{FiniteFlow}: multivariate functional reconstruction
  using finite fields and dataflow graphs}},
  \href{https://doi.org/10.1007/JHEP07(2019)031}{\emph{JHEP} {\bfseries 07}
  (2019) 031} [\href{https://arxiv.org/abs/1905.08019}{{\ttfamily
  1905.08019}}].

\bibitem{Lee:2012cn}
R.N.~Lee, \emph{{Presenting LiteRed: a tool for the Loop InTEgrals REDuction}},
   \href{https://arxiv.org/abs/1212.2685}{{\ttfamily 1212.2685}}.

\bibitem{Lee:2013mka}
R.N.~Lee, \emph{{LiteRed 1.4: a powerful tool for reduction of multiloop
  integrals}}, \href{https://doi.org/10.1088/1742-6596/523/1/012059}{\emph{J.
  Phys. Conf. Ser.} {\bfseries 523} (2014) 012059}
  [\href{https://arxiv.org/abs/1310.1145}{{\ttfamily 1310.1145}}].

\bibitem{Guan:2024byi}
X.~Guan, X.~Liu, Y.-Q.~Ma and W.-H.~Wu, \emph{{Blade: A package for
  block-triangular form improved Feynman integrals decomposition}},
  \href{https://arxiv.org/abs/2405.14621}{{\ttfamily 2405.14621}}.

\bibitem{Gehrmann:1999as}
T.~Gehrmann and E.~Remiddi, \emph{{Differential equations for two-loop
  four-point functions}},
  \href{https://doi.org/10.1016/S0550-3213(00)00223-6}{\emph{Nucl. Phys. B}
  {\bfseries 580} (2000) 485}
  [\href{https://arxiv.org/abs/hep-ph/9912329}{{\ttfamily hep-ph/9912329}}].

\bibitem{Pak:2011xt}
A.~Pak, \emph{{The toolbox of modern multi-loop calculations: novel analytic
  and semi-analytic techniques}},
  \href{https://doi.org/10.1088/1742-6596/368/1/012049}{\emph{J. Phys. Conf.
  Ser.} {\bfseries 368} (2012) 012049}
  [\href{https://arxiv.org/abs/1111.0868}{{\ttfamily 1111.0868}}].

\bibitem{Smirnov:2010hn}
A.V.~Smirnov and A.V.~Petukhov, \emph{{The Number of Master Integrals is
  Finite}}, \href{https://doi.org/10.1007/s11005-010-0450-0}{\emph{Lett. Math.
  Phys.} {\bfseries 97} (2011) 37}
  [\href{https://arxiv.org/abs/1004.4199}{{\ttfamily 1004.4199}}].

\bibitem{Fermat}
R.H.~Lewis, \emph{{Computer Algebra System Fermat}},
  \href{https://home.bway.net/lewis}{https://home.bway.net/lewis}.

\bibitem{Kauers:2008zz}
M.~Kauers, \emph{{Fast Solvers for Dense Linear Systems}},
  \href{https://doi.org/10.1016/j.nuclphysbps.2008.09.111}{\emph{Nucl. Phys. B
  Proc. Suppl.} {\bfseries 183} (2008) 245}.

\bibitem{Kant:2013vta}
P.~Kant, \emph{{Finding linear dependencies in integration-by-parts equations:
  A Monte Carlo approach}},
  \href{https://doi.org/10.1016/j.cpc.2014.01.017}{\emph{Comput. Phys. Commun.}
  {\bfseries 185} (2014) 1473}
  [\href{https://arxiv.org/abs/1309.7287}{{\ttfamily 1309.7287}}].

\bibitem{vonManteuffel:2014ixa}
A.~von Manteuffel and R.M.~Schabinger, \emph{{A novel approach to integration
  by parts reduction}},
  \href{https://doi.org/10.1016/j.physletb.2015.03.029}{\emph{Phys. Lett. B}
  {\bfseries 744} (2015) 101}
  [\href{https://arxiv.org/abs/1406.4513}{{\ttfamily 1406.4513}}].

\bibitem{Peraro:2016wsq}
T.~Peraro, \emph{{Scattering amplitudes over finite fields and multivariate
  functional reconstruction}},
  \href{https://doi.org/10.1007/JHEP12(2016)030}{\emph{JHEP} {\bfseries 12}
  (2016) 030} [\href{https://arxiv.org/abs/1608.01902}{{\ttfamily
  1608.01902}}].

\bibitem{Klappert:2019emp}
J.~Klappert and F.~Lange, \emph{{Reconstructing rational functions with
  FireFly}}, \href{https://doi.org/10.1016/j.cpc.2019.106951}{\emph{Comput.
  Phys. Commun.} {\bfseries 247} (2020) 106951}
  [\href{https://arxiv.org/abs/1904.00009}{{\ttfamily 1904.00009}}].

\bibitem{Klappert:2020aqs}
J.~Klappert, S.Y.~Klein and F.~Lange, \emph{{Interpolation of dense and sparse
  rational functions and other improvements in FireFly}},
  \href{https://doi.org/10.1016/j.cpc.2021.107968}{\emph{Comput. Phys. Commun.}
  {\bfseries 264} (2021) 107968}
  [\href{https://arxiv.org/abs/2004.01463}{{\ttfamily 2004.01463}}].

\bibitem{Fael:2021kyg}
M.~Fael, F.~Lange, K.~Sch\"onwald and M.~Steinhauser, \emph{{A semi-analytic
  method to compute Feynman integrals applied to four-loop corrections to the $
  \overline{\mathrm{MS}} $-pole quark mass relation}},
  \href{https://doi.org/10.1007/JHEP09(2021)152}{\emph{JHEP} {\bfseries 09}
  (2021) 152} [\href{https://arxiv.org/abs/2106.05296}{{\ttfamily
  2106.05296}}].

\bibitem{Maierhofer:2018gpa}
P.~Maierh\"ofer and J.~Usovitsch, \emph{{Kira 1.2 Release Notes}},
  \href{https://arxiv.org/abs/1812.01491}{{\ttfamily 1812.01491}}.

\bibitem{Driesse:2024xad}
M.~Driesse, G.U.~Jakobsen, G.~Mogull, J.~Plefka, B.~Sauer and J.~Usovitsch,
  \emph{{Conservative Black Hole Scattering at Fifth Post-Minkowskian and First
  Self-Force Order}},
  \href{https://doi.org/10.1103/PhysRevLett.132.241402}{\emph{Phys. Rev. Lett.}
  {\bfseries 132} (2024) 241402}
  [\href{https://arxiv.org/abs/2403.07781}{{\ttfamily 2403.07781}}].

\bibitem{Bern:2024adl}
Z.~Bern, E.~Herrmann, R.~Roiban, M.S.~Ruf, A.V.~Smirnov, V.A.~Smirnov et~al.,
  \emph{{Amplitudes, Supersymmetric Black Hole Scattering at
  $\mathcal{O}(G^5)$, and Loop Integration}},
  \href{https://arxiv.org/abs/2406.01554}{{\ttfamily 2406.01554}}.

\bibitem{Fael:2023tcv}
M.~Fael and J.~Usovitsch, \emph{{Third order correction to semileptonic $b\to
  u$ decay: Fermionic contributions}},
  \href{https://doi.org/10.1103/PhysRevD.108.114026}{\emph{Phys. Rev. D}
  {\bfseries 108} (2023) 114026}
  [\href{https://arxiv.org/abs/2310.03685}{{\ttfamily 2310.03685}}].

\bibitem{Fael:2024vko}
M.~Fael, T.~Huber, F.~Lange, J.~M\"uller, K.~Sch\"onwald and M.~Steinhauser,
  \emph{{Heavy-to-light form factors to three loops}},
  \href{https://arxiv.org/abs/2406.08182}{{\ttfamily 2406.08182}}.

\bibitem{Magerya:2022hvj}
V.~Magerya, \emph{{Rational Tracer: a Tool for Faster Rational Function
  Reconstruction}},  \href{https://arxiv.org/abs/2211.03572}{{\ttfamily
  2211.03572}}.

\bibitem{Agarwal:2024jyq}
B.~Agarwal, G.~Heinrich, S.P.~Jones, M.~Kerner, S.Y.~Klein, J.~Lang et~al.,
  \emph{{Two-loop amplitudes for $ t\overline{t}H $ production: the
  quark-initiated N$_{f}$-part}},
  \href{https://doi.org/10.1007/JHEP05(2024)013}{\emph{JHEP} {\bfseries 05}
  (2024) 013} [\href{https://arxiv.org/abs/2402.03301}{{\ttfamily
  2402.03301}}].

\bibitem{Chawdhry:2023yyx}
H.A.~Chawdhry, \emph{{p-adic reconstruction of rational functions in multi-loop
  amplitudes}},  \href{https://arxiv.org/abs/2312.03672}{{\ttfamily
  2312.03672}}.

\bibitem{Liu:2023cgs}
X.~Liu, \emph{{Reconstruction of rational functions made simple}},
  \href{https://doi.org/10.1016/j.physletb.2024.138491}{\emph{Phys. Lett. B}
  {\bfseries 850} (2024) 138491}
  [\href{https://arxiv.org/abs/2306.12262}{{\ttfamily 2306.12262}}].

\bibitem{Mokrov:2023vva}
K.~Mokrov, A.~Smirnov and M.~Zeng, \emph{{Rational Function Simplification for
  Integration-by-Parts Reduction and Beyond}},
  \href{https://arxiv.org/abs/2304.13418}{{\ttfamily 2304.13418}}.

\bibitem{Wu:2024paw}
Z.~Wu and Y.~Zhang, \emph{{A new method for finding more symmetry relations of
  Feynman integrals}},  \href{https://arxiv.org/abs/2406.20016}{{\ttfamily
  2406.20016}}.

\bibitem{Vermaseren:1994je}
J.A.M.~Vermaseren, \emph{{Axodraw}},
  \href{https://doi.org/10.1016/0010-4655(94)90034-5}{\emph{Comput. Phys.
  Commun.} {\bfseries 83} (1994) 45}.

\bibitem{Binosi:2003yf}
D.~Binosi and L.~Theu{\ss}l, \emph{{JaxoDraw: A Graphical user interface for
  drawing Feynman diagrams}},
  \href{https://doi.org/10.1016/j.cpc.2004.05.001}{\emph{Comput. Phys. Commun.}
  {\bfseries 161} (2004) 76}
  [\href{https://arxiv.org/abs/hep-ph/0309015}{{\ttfamily hep-ph/0309015}}].

\bibitem{Harlander:2020cyh}
R.V.~Harlander, S.Y.~Klein and M.~Lipp, \emph{{FeynGame}},
  \href{https://doi.org/10.1016/j.cpc.2020.107465}{\emph{Comput. Phys. Commun.}
  {\bfseries 256} (2020) 107465}
  [\href{https://arxiv.org/abs/2003.00896}{{\ttfamily 2003.00896}}].

\end{thebibliography}\endgroup

\end{document}